\documentclass[11pt,a4paper]{article}
\usepackage{amsmath}
\usepackage{newtxtext,newtxmath}
\usepackage{graphicx}
\usepackage{epstopdf}
\usepackage{float}
\usepackage{comment}
\usepackage{url}
\usepackage{authblk}
\usepackage{fancyhdr}
\usepackage{braket}
\usepackage{slashed}

\begin{document}

\title{Is Anomaly Transferred thorough Multi-loop Process?}
\author{Toshiki Nakawaki}
\affil{\textit{Department of Physics, Kyoto University, Kyoto 606-8502, Japan}}
\date{}
\maketitle
\thispagestyle{fancy}
\rhead{\textrm{KUNS}-2726}

\begin{abstract}
We investigated the multi-loop anomaly transfer to QCD sector from another one and its ability to solve strong CP problem. 
If the anomalous symmetry is spontaneously broken, its Nambu-Goldstone (NG) boson can couple to the QCD $G\tilde{G}$ through this transfer effect 
and behave as an axion. 
In our result, such a particle acquire mass pertabatively in fact, 
and really massless mode doesn't have couplings with $G\tilde{G}$. 
Consequently, this pseudo-NG boson is found not to solve the problem.
\end{abstract}

\subsection*{Introduction}

The strong CP problem and the smallness of neutrino mass are both serious \textit{problems}, in view of naturalness, included in Standard Model. 
The former is mainly explained by axion models with Pecci-Quinn mechanism~\cite{pq}, 
while the latter is often attributed to seesaw mechanisms~\cite{seesaw}. 
There are also some works handling these problems comprehensively: 
Shin pointed out that the scales of the two mechanisms above should be the same order, 
and built a model with one heavy quark and one complex scaler, 
whose VEV gives neutrino Majorana mass, while the phase behaves as axion~\cite{shin}. 
This axion is also the NG boson of broken $U(1)_l$ symmetry, 
which already had been named "Majoron" by Chikashige, Mohapatra and Peccei~\cite{majoron}, 
and similar "Majoron $=$ Axion" models followed this work~\cite{ma,ohata}.
These models requires additional left- and right-handed quarks with opposite $U(1)_l$ charge 
in order to identify this symmetry as Pecci-Quinn symmetry $U(1)_\mathrm{PQ}$, 
just as the KSVZ axion model~\cite{k,svz}. 

On the other hand, Latosi\'nski, Meissner and Nicolai~\cite{nicolai} proposed a model of this type 
without any new field except for one scalar which generates Majoron. 
In that model, 3-loop Majoron-gluon-gluon couplings are discussed to induce effective axion-like coupling, 
but such contributions seem to be forbidden by Adler-Bardeen theorem. 

We investigated similar situation in general, 
and found that even if there is such a multi-loop anomaly transfer effect, 
the NG boson corresponding to originally anomalous symmetry doesn't change the QCD vacuum angle and cannot work as axion. 
In this paper, we first present an example toy model, and then give general discussion. 

\subsection*{Toy model example}
Consider a model consisting of massless QED, 1-flavor massive QCD and complex scalar field $S$ which couples to electron bilinear: 
\begin{align}
\mathcal{L}&=\mathcal{L}_0+\mathcal{L}_S,\\
\mathcal{L}_0&=-\frac{1}{4}F_{\mu\nu}F^{\mu\nu}-\frac{1}{4}G_{\mu\nu}G^{\mu\nu}
+i\bar{e}\slashed{D}e+i\bar{\mathrm{q}}\slashed{D}\mathrm{q}-m\bar{\mathrm{q}}\mathrm{q},\\
\mathcal{L}_S&=-|\partial_\mu S|^2-V(|S|^2)-y(S\bar{e}P_\mathrm{L}e+\mathrm{h.c.}).
\end{align}
Scalar potential $V$ is assumed to have continuous minima $|S|=M$. 
Because of the symmetry about following chiral transformation (let's call it $U(1)_{l\mathrm{A}}$):
\begin{align}
\begin{cases}
e\to e^{i\alpha\gamma_5}e\\
S\to e^{2i\alpha}S
\end{cases},
\end{align}
the phase of $S$ becomes Nambu-Goldstone boson of this symmetry. 
Corresponding Noether current is
\begin{align}
J_{l\mathrm{A}}^\mu=\bar{e}\gamma^\mu\gamma_5e+2iS^\dagger\overleftrightarrow{\partial}^\mu S.
\end{align}
The NG boson $a$ is normalized as 
\begin{align}
S\simeq M\exp\left( i\frac{a}{\sqrt{2}M} \right)
\end{align}
and couples to $J_{l\mathrm{A}}^\mu$, 

In this model, $a$ couples to quark chiral current $J_{\mathrm{qA}}^\mu$:
\begin{align}
-i\mathcal{M}_{a\mathrm{qq}}=&
\begin{array}{c}
\includegraphics[width=6cm]{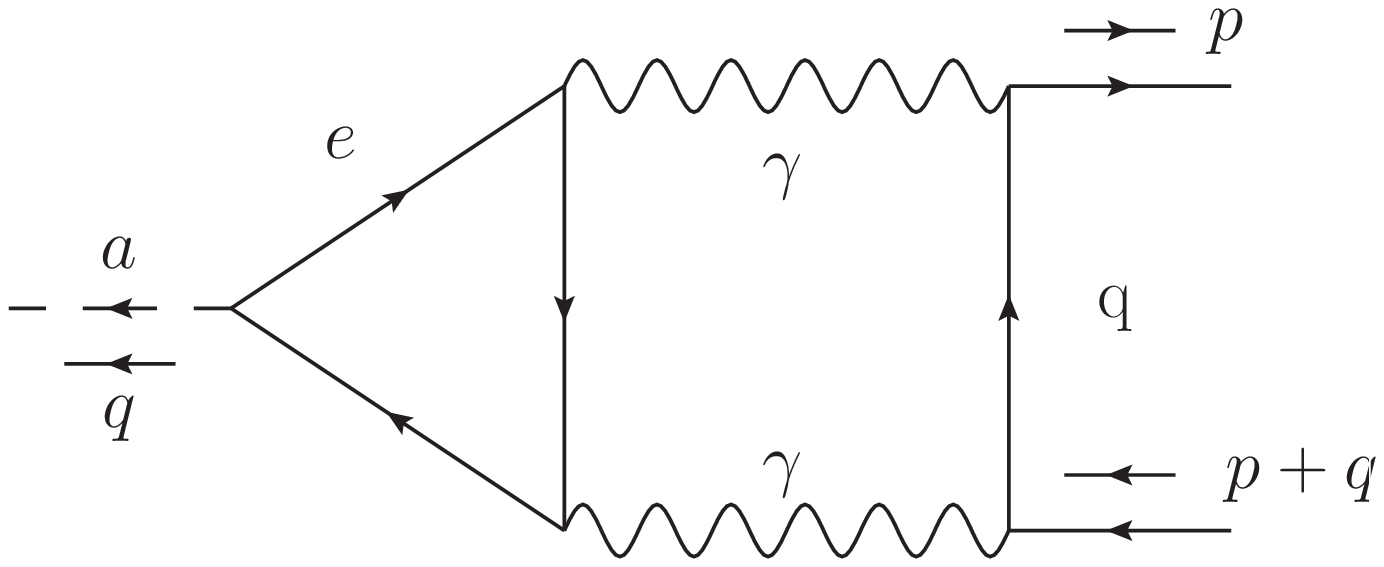}
\end{array}
\notag\\
=&\;-6\sqrt{2}iye^2e_\mathrm{q}^2(\slashed{q}\gamma_5)\notag\\
&\quad\times\int\!\frac{d^4k_1}{(2\pi)^4}\frac{d^4k_2}{(2\pi)^4}
\frac{m_e}{(k_1^2-m_e)^2\{(k_1+k_2)^2-m_e^2\}k_2^2(k_2^2-m^2)}\notag\\
(=&\;c_{a\mathrm{qq}}\slashed{q}\gamma_5)\\
\Rightarrow\qquad\mathcal{L}_{a\mathrm{qq,eff}}=&\;-c_{a\mathrm{qq}}\partial_\mu a\bar{\mathrm{q}}\gamma^\mu\gamma_5\mathrm{q}.
\end{align}
Here, $e_\mathrm{q}$ is quark electric charge and $m_e=yM$. 
This means that the current operator $J_{l\mathrm{A}}^\mu$ mixes with $J_{\mathrm{qA}}^\mu$ through higher loops, 
and then $a$ couples with $G_{\mu\nu}\tilde{G}^{\mu\nu}$ through its divergence. 
However, this mixed current doesn't conserve (even without considering anomaly) because of quark mass $m\neq 0$, 
and the corresponding NG boson $a$ \textit{acquire a mass}. 
In order to specify the (pertabatively-at-least) massless NG mode (which must uniquely exist because of Goldstone theorem), 
one have to find conserved current (except for anomaly effect). The answer is 
\begin{align}
\tilde{J}_{l\mathrm{A}}=J_{l\mathrm{A}}-4\sqrt{2}Mc_{a\mathrm{qq}}J_{\mathrm{qA}}.
\end{align}
Therefore the orthogonal linear combination $4\sqrt{2}Mc_{a\mathrm{qq}}a+\eta$ 
($\eta$ is massive NG mode corresponding to explicitly broken 'flavor' $U(1)_\mathrm{A}$ symmetry) 
becomes massless mode. 
The corresponding current $\tilde{J}_{l\mathrm{A}}$ doesn't couple to quark current, so this combination never feels QCD vacuum potential.

\subsection*{General Discussion}
Such superficial transfer of anomaly occurs in arbitrary models of fermions with spontaniously broken chiral symmetry 
if they are connected with massive colored quarks by some gauge interaction other than gluon. 
For convenience, we restrict our attention on 1-flavor QCD with quark mass $m$, but generalization is trivial. 
With massive quarks, this model has $\theta$-vacua structure.

\begin{figure}
\begin{center}
\includegraphics[height=3cm]{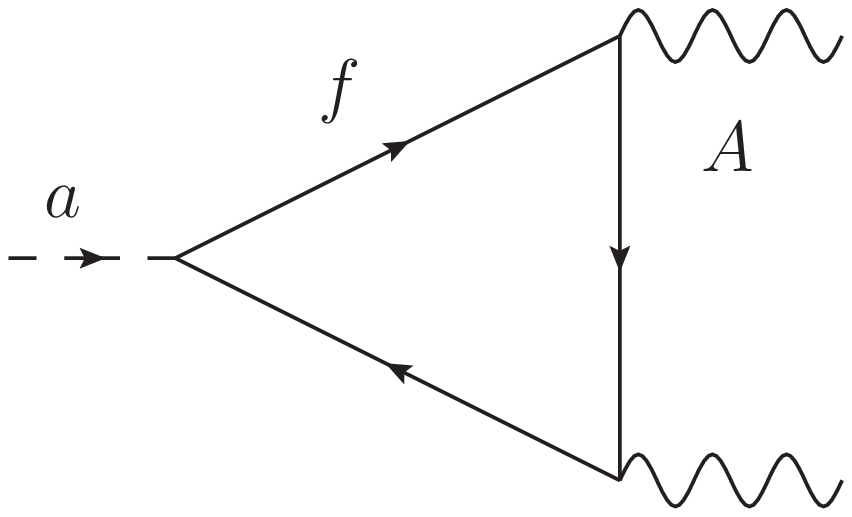}\\
\includegraphics[height=3cm]{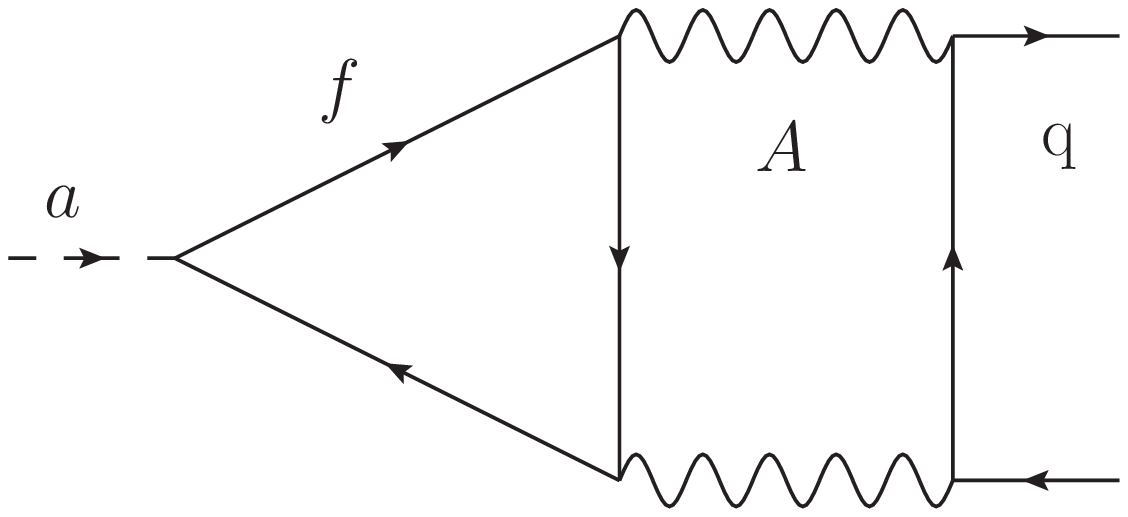}\\
\includegraphics[height=3cm]{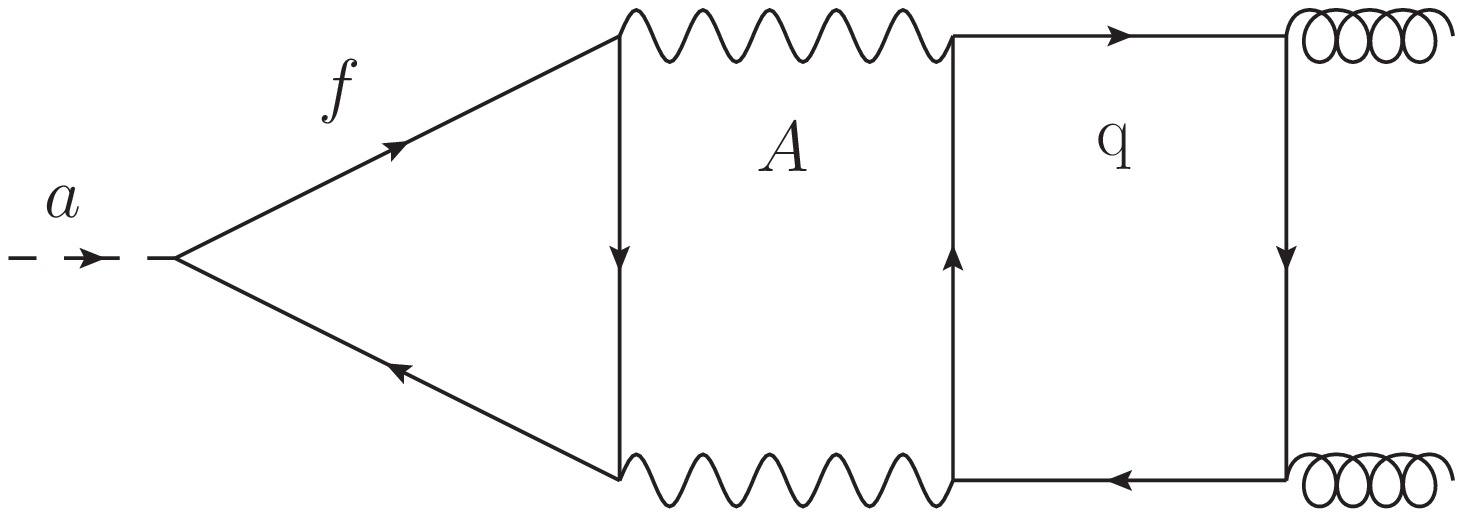}
\caption{Various processes including $a$-$F\tilde{F}$ anomalous triangle. $f$ represents a fermion in lepton sector and $A$ is a mediating gauge field. \label{a_graphs}}
\end{center}
\end{figure}

The first sector is intended to represent leptonic one, so let's extend the use of this term to general case. 
This sector has a (pertabatively) massless mode $a$ as usual NG boson, 
and this couples to CP violating combination of gauge field strength $F_{\mu\nu}\tilde{F}^{\mu\nu}$ through anomalous triangle in figure \ref{a_graphs}. 
As shown togather, with QCD, this couples to quark current and even to gluons. 
Especially, the third diagram in figure \ref{a_graphs} includes, because of anomalous subdiagram, 
partial amplitude proportional to $\epsilon^{\mu\nu\rho\sigma}$, introducing $a$-$G\tilde{G}$ effective coupling. 

\begin{figure}
\begin{center}
\includegraphics[height=3cm]{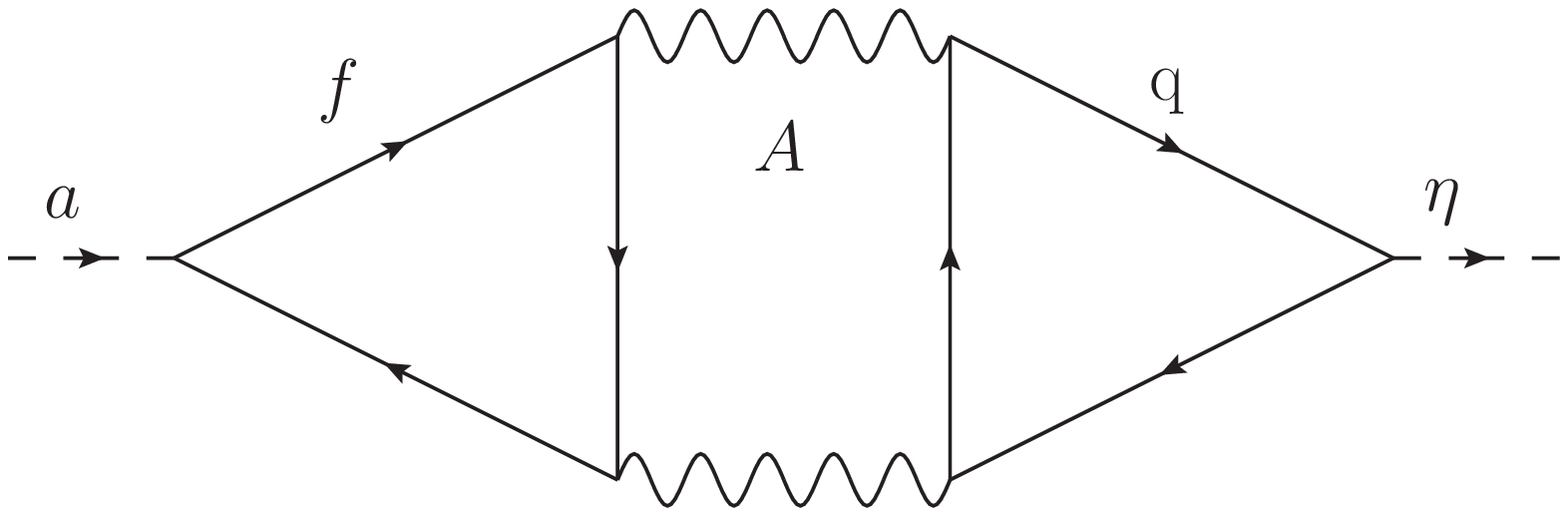}
\caption{An example of diagrams that enable $a$-$\eta$ mixing. \label{a_eta_graphs}}
\end{center}
\end{figure}

On the other hand, in $m\to 0$ limit QCD sector also has chiral $U(1)$ symmetry on quarks, 
and for $m\neq 0$, its 'NG' boson $\eta$ exists as massive mode. 
This also couples anomalous triangle of quarks and both $F\tilde{F}$ and $G\tilde{G}$ for gluon field strength $G$. 
Such couplings enables $a$ and $\eta$ to mix with each other through diagrams like one in figure \ref{a_eta_graphs}.

After integrating out all fields except for the two scalars, their effective lagrangian has a kinetic term as follows:
\begin{align}
\mathcal{L}_{a\eta,\mathrm{kin}}=\frac{1}{2}\partial_\mu a\partial^\mu a+\frac{1}{2}\partial_\mu\eta\partial^\mu\eta+C\partial_\mu a\partial^\mu\eta.\label{effkin}
\end{align}
The third term comes from $a$-$\eta$ mixing diagrams, 
which are always bilinear to momenta of them and induce only kinetic mixing. 

\begin{figure}
\begin{center}
\includegraphics[width=8cm]{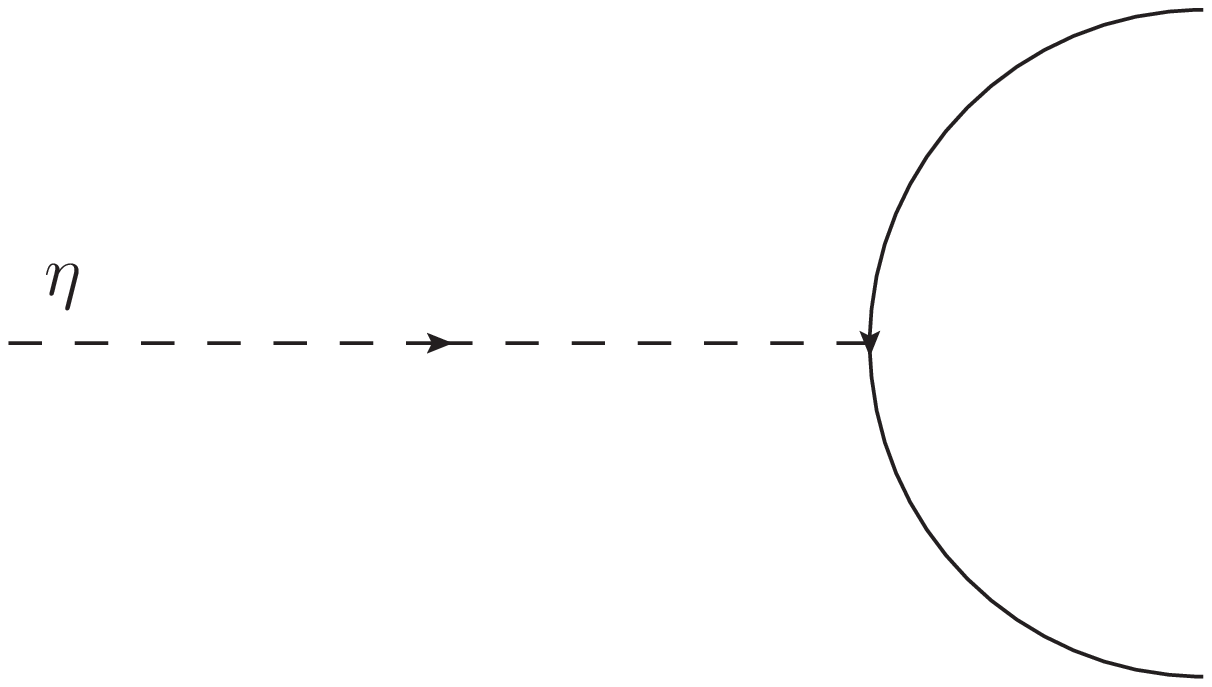}
\includegraphics[width=8cm]{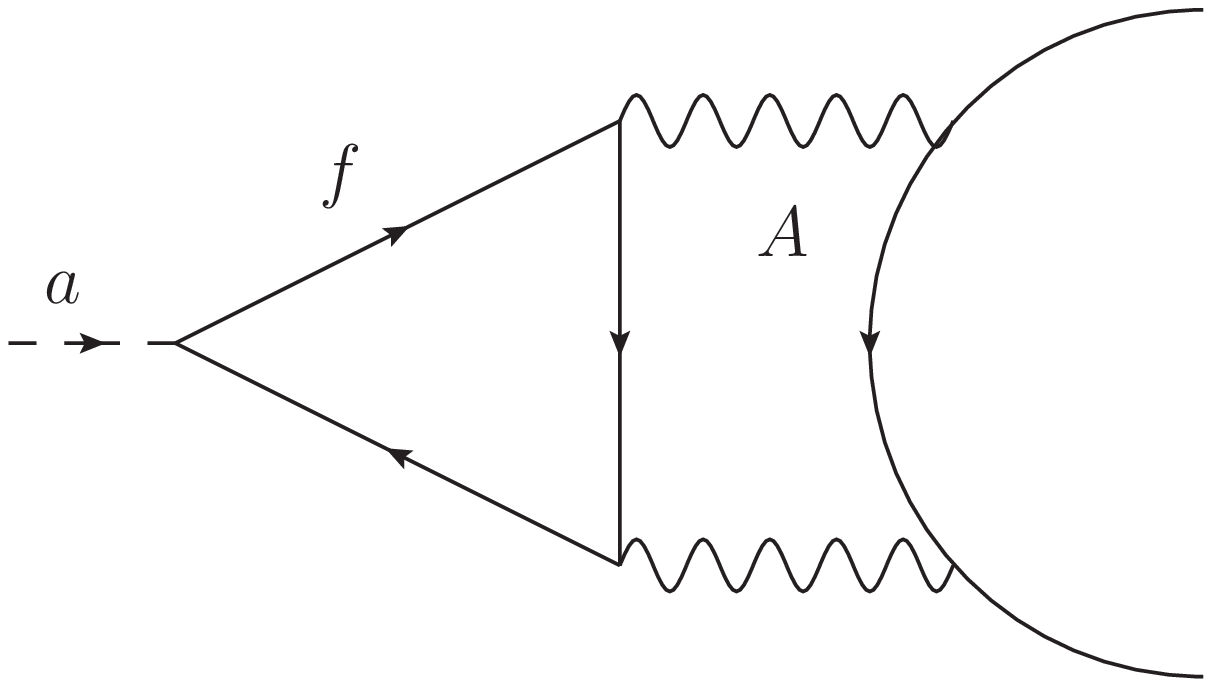}
\caption{Subdiagrams composing $\eta$ and $a$ 2 point amplitude generating their masses. \label{aqq_etaqq}}
\end{center}
\end{figure}

Mass term of the effective lagrangian comes from two effects: 
$\theta$-vacua of gauge fields and quark mass. 
The former is generated by gluons, and also mediating fields $A$, if their corresponding gauge group has $SU(2)$ subgroup. 
However, we will take only potential from gluons into account and neglect the vacuum structure of $A$'s. 
Then, this gives mass to $\eta$ through anomalous q's triangle, and to $a$ through the third diagram in figure \ref{a_graphs}. 
The latter generates $\eta$ mass directly by quark loop, 
and one or both of the two external legs can be replaced $a$ with the second two-loop diagram in figure \ref{a_graphs}. 
Therefore quark mass induces their mass term by 4 diagrams generated by connecting any 2 of 2 subdiagrams shown in figure \ref{aqq_etaqq}.
In consiquense, this contribution to mass term lagrangian must depend fields in a form of a certain linear combination $\eta+ra$, 
where $r$ is the ratio of two diagrams in figure \ref{aqq_etaqq}. 
This is also the case for contribution of $\theta$-vacua 
because both $a$ and $\eta$ couples to $J_{\mathrm{qA}}^\mu$ through diagrams in figure \ref{aqq_etaqq}. 
Therefore complete effective mass lagrangian is 
\begin{align}
\mathcal{L}_{a\eta,\mathrm{mass}}=\frac{\mu^2}{2}(\eta+ra)^2.\label{effmass}
\end{align}

The total effective lagrangian (\ref{effkin})$+$(\ref{effmass}) is harmonic oscilator-like, and able to diagonalize with a certain two field basis: 
redefine $\eta$ as $\eta+ra$, diagonalize the kinetic terms and rotate mass terms with orthogonal matrix. 
Because original mass matrix is not rankful, there exists one free field after this diagonalization. 
This means that this mode is pertabatively massless and doesn't feel QCD $\theta$-vacua. 
The other field basis, which couples to $\theta$-vacua, acquires mass pertabatively, 
and therefore neither of them is capable of setting vacuum angle $\theta$ small. 

\subsection*{Conclusion}
We revealed that axion models depending on anomaly tranfer doesn't work in general case. 
Actually anomaly does transfer from its original symmetry to others, 
but as far as colored quarks have canonical mass, this transfer to QCD sector cannot solve the strong CP problem. 
This is also the case for QCD with quark mass generated in some spontaneous symmetry breaking, 
so there is no problem in replacing the QCD sector with quark sector of the Standard Model 
if the Higgs scalar has nothing to do with the chiral symmetry of lepton sector. 

This result tells that axions must couple with quarks in tree level, or they don't move QCD vacuum angle. 
So in order to assume Majoron as axion, it have to couple with additional colored quarks. 

\subsection*{Acknowledgment}
The author was helped very much by Hikaru Kawai, Koji Tsumura and Koichi Yoshioka throughout this work. 
Yu Hamada and Sho Higuchi are also thanked for their contribution with ideas and calculations.

\bibliographystyle{unsrt}
\bibliography{bibdata}

\end{document}